\documentclass[12pt]{iopart}

\usepackage{cite}
\usepackage{graphicx}
\usepackage{units}
\usepackage[colorlinks,
breaklinks,
citecolor=blue,
linkcolor=blue,
]{hyperref}
\begin{document}

\title[Optical-pumping enantio-conversion of chiral mixtures]{Optical-pumping enantio-conversion of chiral mixtures in presence of tunneling between chiral states}

\author{Fen Zou\textsuperscript{1,2,3}, Chong Ye\textsuperscript{4} and Yong Li\textsuperscript{1,*}}

\address{\textsuperscript{1} Center for Theoretical Physics, Hainan University, Haikou 570228, China}
\address{\textsuperscript{2} Beijing Computational Science Research Center, Beijing 100193, China}
\address{\textsuperscript{3} Synergetic Innovation Center for Quantum Effects and Applications, Hunan Normal University, Changsha 410081, China}
\address{\textsuperscript{4} Beijing Key Laboratory of Nanophotonics and Ultrafine Optoelectronic Systems, School of Physics, Beijing Institute of Technology, Beijing 100081, China}
\address{\textsuperscript{*} Author to whom any correspondence should be addressed.}
\ead{yongli@hainanu.edu.cn}

\vspace{10pt}

\begin{abstract}
Enantio-conversion of chiral mixtures, converting the mixtures composed of left- and right-handed chiral molecules into the homochiral ensembles, has become an important research topic in chemical and biological fields. In previous studies on enantio-conversion, the tunneling interaction between the left- and right-handed chiral states was often neglected. However, for certain chiral molecules, this tunneling interaction is significant and cannot be ignored. Here we propose a scheme for enantio-conversion of chiral mixtures through optical pumping based on a four-level model of chiral molecules, comprising two chiral ground states and two achiral excited states, with a tunneling interaction between the chiral states. Under one-photon large detuning and two-photon resonance conditions, one of the achiral excited states is eliminated adiabatically. By well designing the detuning and coupling strengths of the electromagnetic fields, the tunneling interaction between two chiral states and the interaction between one of the chiral states and the remaining achiral excited state can be eliminated. Consequently, one chiral state remains unchanged, while the other can be excited to an achiral excited state, establishing chiral-state-selective excitations. By numerically calculating the populations of two chiral ground states and the enantiomeric excess, we observe that high-efficiency enantio-conversion is achieved under the combined effects of system dissipation and chiral-state-selective excitations.
\end{abstract}

\vspace{2pc}

\submitto{\NJP}

\maketitle

\section{Introduction}
Many molecules exhibit chirality, existing in left- and right-handed forms. The left-handed chiral molecule cannot be superimposed with its mirror image (i.e., the right-handed one) via translation and rotation~\cite{Woolley1976Quantum}. These left- and right-handed molecules are referred to as enantiomers. Enantiomers share almost the same physical properties, such as boiling points, melting points, and densities, yet they possess divergent biological activities and functions~\cite{mezey1991global}. For example, one enantiomeric form is beneficial in designing pharmaceuticals, while the other is useless or even harmful. In the field of medicine, acquiring homochiral medicines that offer benefits to humans is of paramount importance. Consequently, enantio-discrimination~\cite{Jia2011Probing,Hirota2012Triple,Yachmenev2016Detecting,Lehmann2018Influence,Chen2020Enantio,
Xu2020Enantiomeric,Kang2020Effective,Ye2021Entanglement,Chen2021arXivEnantio,cai2022Enantiodetection,zou2022Enantiodiscrimination,ye2023SingleShot,Mu2023Machine,ye2023Phasematched}, spatial enantio-separation~\cite{Li2007Generalized,Li2010Theory,Jacob2012Effect,Eilam2013Spatial,Bradshaw2015Laser,Liu2021Spatial,Suzuki2019Stern,
Milner2019Controlled}, enantio-specific state transfer~\cite{Kral2001Cyclic,Li2008Dynamic,Jia2010Distinguishing,Leibscher2019Principles,Ye2019Effective,
Vitanov2019Highly,Wu2019Robust,Wu2020Two,Wu2020Discrimination,Torosov2020Efficient,Torosov2020Chiral,Guo2022Cyclic}, and enantio-conversion~\cite{shapiro2000Coherently,gerbasi2001Theory,brumer2001Principles,kral2003TwoStep,thanopulos2003Theory,
frishman2004Optical,ye2021Improved,ye2020Fast,ye2021Enantioconversion} of chiral mixtures have become crucial research topics in the chemical and biological fields.

In the past two decades, various theoretical schemes on enantio-discrimination~\cite{Jia2011Probing,Hirota2012Triple,Yachmenev2016Detecting,Lehmann2018Influence,Chen2020Enantio,
Xu2020Enantiomeric,Kang2020Effective,Ye2021Entanglement,Chen2021arXivEnantio,cai2022Enantiodetection,zou2022Enantiodiscrimination,ye2023SingleShot,Mu2023Machine,ye2023Phasematched}, spatial enantio-separation~\cite{Li2007Generalized,Li2010Theory,Jacob2012Effect,Eilam2013Spatial,Bradshaw2015Laser,Liu2021Spatial,Suzuki2019Stern,
Milner2019Controlled}, and enantio-specific state transfer~\cite{Kral2001Cyclic,Li2008Dynamic,Jia2010Distinguishing,Leibscher2019Principles,Ye2019Effective,
Vitanov2019Highly,Wu2019Robust,Wu2020Two,Wu2020Discrimination,Torosov2020Efficient,Torosov2020Chiral} have been proposed based on the three-level $\Delta$-type model of chiral molecules. This model is composed of three electromagnetic fields coupled respectively to three related electric-dipole transitions of chiral molecules. By taking advantage of the property of the overall phase-difference $\pi$ of the three Rabi frequencies for the two enantiomers, enantio-discrimination~\cite{Jia2011Probing,Hirota2012Triple,Yachmenev2016Detecting,Lehmann2018Influence,Chen2020Enantio,
Xu2020Enantiomeric,Kang2020Effective,Ye2021Entanglement,Chen2021arXivEnantio,cai2022Enantiodetection,zou2022Enantiodiscrimination,ye2023SingleShot,Mu2023Machine,ye2023Phasematched}, spatial enantio-separation~\cite{Li2007Generalized,Li2010Theory,Jacob2012Effect,Eilam2013Spatial,Bradshaw2015Laser,Liu2021Spatial,Suzuki2019Stern,
Milner2019Controlled}, and enantio-specific state transfer~\cite{Kral2001Cyclic,Li2008Dynamic,Jia2010Distinguishing,Leibscher2019Principles,Ye2019Effective,
Vitanov2019Highly,Wu2019Robust,Wu2020Two,Wu2020Discrimination,Torosov2020Efficient,Torosov2020Chiral} of chiral mixtures can be achieved. Particularly, the experiments on enantio-discrimination~\cite{Patterson2013Enantiomer,Patterson2013Sensitive,
Patterson2014New,Shubert2014Identifying,Shubert2015Rotational,Shubert2016Chiral,Lobsiger2015Molecular} and enantio-specific state transfer~\cite{Eibenberger2017Enantiomer,Perez2017Coherent,Lee2022Quantitative} have been achieved based on the three-level $\Delta$-type model of chiral molecules. We would like to note that a similar energy transfer during propagating between four-wave-mixing and six-wave-mixing signals has been theoretically and experimentally investigated via atomic coherence in a four-level inverted-Y atomic system, falling in the electromagnetically induced transparency window~\cite{Zhang2008Efficient,Zhang2009Temporal}.

Recently, utilizing the four-level (or five-level) double-$\Delta$ model~\cite{shapiro2000Coherently,gerbasi2001Theory,brumer2001Principles,kral2003TwoStep,thanopulos2003Theory,
frishman2004Optical,ye2021Improved,ye2020Fast,ye2021Enantioconversion}, some theoretical methods for enantio-conversion of chiral mixtures have been proposed, such as the laser-distillation method~\cite{shapiro2000Coherently,gerbasi2001Theory,brumer2001Principles,kral2003TwoStep,thanopulos2003Theory,
frishman2004Optical,ye2021Improved}, coherent-operation method~\cite{ye2020Fast}, and optical-pumping method~\cite{ye2021Enantioconversion}. Enantio-conversion aims to convert chiral mixtures to enantiopure samples with the desired chirality. The laser-distillation method~\cite{shapiro2000Coherently,gerbasi2001Theory,brumer2001Principles,kral2003TwoStep,thanopulos2003Theory,
frishman2004Optical,ye2021Improved} achieves enantio-conversion by repeating a pair of excitation and relaxation steps, but it is time-consuming and the related efficiency of achieving enantio-conversion is not high. To achieve highly efficient enantio-conversion of chiral mixtures, Ye \textit{et al}. proposed coherent-operation~\cite{ye2020Fast} and optical-pumping~\cite{ye2021Enantioconversion} methods. Comparing with the laser-distillation method~\cite{shapiro2000Coherently,gerbasi2001Theory,brumer2001Principles,kral2003TwoStep,thanopulos2003Theory,
frishman2004Optical,ye2021Improved}, the purely coherent-operation method~\cite{ye2020Fast} can shorten the required time by three orders of magnitude and achieve high-efficiency enantio-conversion, while the optical-pumping method~\cite{ye2021Enantioconversion} has no requirement of precise control of pulse areas or pulse shapes. In the optical-pumping method, the decoherence from the achiral excited states to the two chiral ground states plays a beneficial role in achieving enantio-conversion.

Note that in previous enantio-conversion works~\cite{shapiro2000Coherently,gerbasi2001Theory,brumer2001Principles,kral2003TwoStep,thanopulos2003Theory,
frishman2004Optical,ye2021Improved,ye2020Fast,ye2021Enantioconversion}, the tunneling interaction between left- and right-handed chiral states was often neglected since the tunneling effect is usually considered to be sufficiently weak. However, for certain chiral molecules, the magnitude of the tunneling interaction strength may be comparable to the coupling strength, e.g. the tunneling can occur within $33\,\mathrm{ms}-3.3\,\mu\mathrm{s}$ for small chiral molecules (like $\mathrm{D}_{2}\mathrm{S}_{2}$)~\cite{thanopulos2003Theory,Jeffrey1995American}, namely, the tunneling interaction between two chiral ground states can not be ignored. It is therefore natural to ask the question: how to achieve high-efficiency enantio-conversion of chiral mixtures in this case?

In this paper, we propose achieving high-efficiency enantio-conversion of chiral mixtures via optical pumping based on a four-level model of chiral molecules, composed of two chiral ground states and two achiral excited states. Compared with the previous four-level double-$\Delta$ model~\cite{shapiro2000Coherently,gerbasi2001Theory,brumer2001Principles,kral2003TwoStep,thanopulos2003Theory,
frishman2004Optical,ye2021Improved,ye2020Fast}, additional tunneling interaction between the two chiral ground states is introduced here~\cite{grishanin1999Photoinduced,bychkov2001Laser,bychkov2002Laser,zhdanov2007Absolute,zhdanov2010Coherent,
quack2021Chapter,stickler2021Enantiomer,sun2023Inducinga}. Under the condition of the large detuning between the two chiral ground states and the symmetric achiral excited state, as well as the two-photon resonance between the two chiral ground states and the asymmetric achiral excited state, the symmetric achiral excited state can be eliminated adiabatically. In this case, by well designing the detuning and coupling strengths of the electromagnetic fields, the tunneling interaction between two chiral ground states and the interaction between the left-handed ground state and the asymmetric achiral excited state can be counteracted. Therefore, the left-handed chiral ground state remains undisturbed, while the right-handed one is excited to the asymmetric achiral excited state, establishing chiral-state-selective excitations. Meanwhile, this achiral excited state relaxes to two chiral ground states due to the system dissipation and thus the enantio-conversion of chiral mixtures can be realized in the steady state. To assess the efficiency of the enantio-conversion of chiral mixtures, we calculate numerically the populations of two chiral ground states and the enantiomeric excess of the chiral ground state. In addition, we analyze the effect of the system parameters (e.g. the detuning and the coupling strength) on enantio-conversion of chiral mixtures.

\section{Model and Hamiltonian\label{modelsec}}
\begin{figure}
\center
\includegraphics[width=0.6 \textwidth]{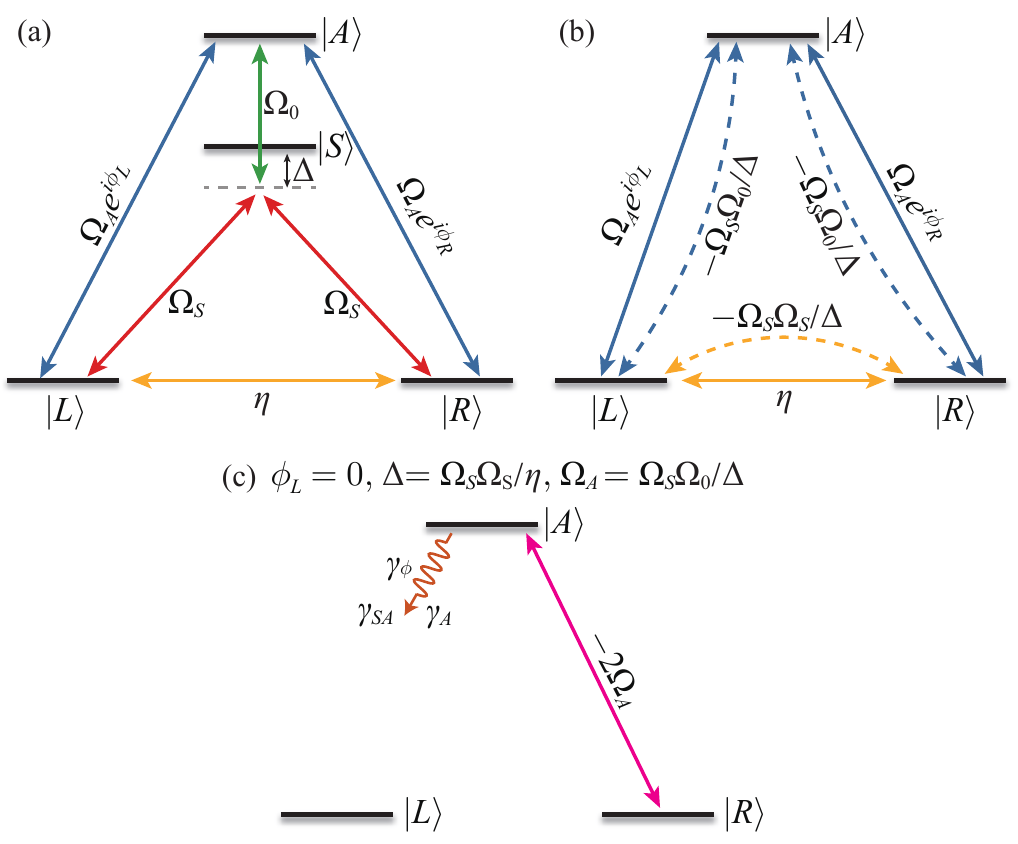}
\caption{(a) The schematic diagram of the four-level model of chiral molecules. Here $\vert L\rangle$ and $\vert R\rangle$ are, respectively, the degenerated left- and right-handed chiral ground states, while $\vert S\rangle$ and $\vert A\rangle$ are, respectively, the symmetric and asymmetric achiral excited states. Three electromagnetic fields with frequencies $\omega_{2}$, $\omega_{0}$, and $\omega_{1}$ are applied to couple the four-level model in $\Delta$-type substructures of $\vert Q\rangle\leftrightarrow\vert A\rangle\leftrightarrow\vert S\rangle\leftrightarrow\vert Q\rangle$ with $Q=L,R$ under three-photon resonance conditions, where the corresponding coupling strengths are $\Omega_{A}$, $\Omega_{0}$, and $\Omega_{S}$. $\phi_{L}$ and $\phi_{R}$ are the overall phases of the two $\Delta$-type substructures with $\phi_{R}=\phi_{L}+\pi$. In addition, there is a tunneling interaction between $\vert L\rangle$ and $\vert R\rangle$, with $\eta$ being the tunneling strength. (b) The four-level model is simplified to the effective three-level model by eliminating adiabatically the symmetric achiral excited state $\vert S\rangle$ in the large-detuning region. Here the blue dashed lines represent the indirect interaction between $\vert Q\rangle$ and $\vert A\rangle$, with a coupling strength of $-\Omega_{S}\Omega_{0}/\Delta$, induced by the two-photon processes $\vert Q\rangle\leftrightarrow\vert S\rangle\leftrightarrow\vert A\rangle$. Similarly, the orange dashed line indicates the indirect tunneling interaction between $\vert L\rangle$ and $\vert R\rangle$ with coupling strength $-\Omega_{S}^{2}/\Delta$ that is induced by the process $\vert L\rangle\leftrightarrow\vert S\rangle\leftrightarrow\vert R\rangle$. (c) Under the condition of $\phi_{L}=0$ (that means $\phi_{R}=\pi$), $\Delta=\Omega_{S}^{2}/\eta$, and $\Omega_{A}=\Omega_{S}\Omega_{0}/\Delta$, the left-handed chiral ground state $\vert L\rangle$ is decoupled to $\vert R\rangle$ and $\vert A\rangle$.}
\label{Fig1}
\end{figure}

As shown in Fig.~\ref{Fig1}(a), we consider a four-level model of chiral molecules consisting of two degenerated chiral ground states (the left- and right-handed chiral ground states $\vert L\rangle$ and $\vert R\rangle$) and two achiral excited states (the symmetric and asymmetric achiral excited states $\vert S\rangle$ and $\vert A\rangle$).
The energies of the four states are $\hbar\omega_{A}>\hbar\omega_{S}>\hbar\omega_{L}=\hbar\omega_{R}=0$. Here, the reason  for the degeneracy of the two chiral ground states is that the tiny parity violating energy difference caused by the fundamental weak force is negligible~\cite{Quack2008High}.

In this work, we mainly focus on the case that all of the four states of the chiral molecules under consideration are in the electronic ground state (with different vibrational states) and the barrier of the double-well potential is relatively low. For instance, HSOH molecules~\cite{Yurchenko2009An} can be considered as an example to realize the four-level model of chiral molecules under consideration. Here we take the vibrational sublevels of $\vert L\rangle$ and $\vert R\rangle$ as the vibrational ground states in the left and right wells of the double-well potential, respectively. Obviously, $\vert L\rangle$ and $\vert R\rangle$ are chiral states. The vibrational sublevels of $\vert S\rangle$ and $\vert A\rangle$ are the symmetric and asymmetric vibrational excited states (e.g. near or beyond the barrier of the double-well potential) with a distinguishable vibrational energy difference. When the rotational sublevels of the four states as well as the polarizations and frequencies of three electromagnetic fields are well designed, the transitions out of our working four-level model can be ignored. Specifically, we choose the rotational sublevels of $\vert L\rangle$, $\vert R\rangle$, $\vert S\rangle$, and $\vert A\rangle$ as $\vert J_{\mathrm{k}_{\mathrm{a}}\mathrm{k}_{\mathrm{c}}\mathrm{M}}=0_{000}\rangle$, $\vert0_{000}\rangle$, $\vert1_{010}\rangle$, and $(\vert1_{101}\rangle+\vert1_{10-1}\rangle)/\sqrt{2}$~\cite{Leibscher2019Principles,Ye2018Real}, respectively. Here $\vert J_{\mathrm{k}_{\mathrm{a}}\mathrm{k}_{\mathrm{c}}\mathrm{M}}\rangle$ are eigenstates of asymmetric-top molecules~\cite{Leibscher2019Principles,Ye2018Real}. The three electromagnetic fields associated with the electric-dipole transitions $\vert Q\rangle\leftrightarrow\vert S\rangle\leftrightarrow\vert A\rangle\leftrightarrow\vert Q\rangle$ are, respectively, Z-polarized, X-polarized, and Y-polarized fields. Since the wave vectors ($\vec{k}_{A}$, $\vec{k}_{0}$, and $\vec{k}_{S}$) of the three electromagnetic fields cannot be parallel, the finite size of the sample inevitably produces the phase-mismatching problem. To ensure that the effect of the phase mismatching is negligible, the characteristic length of the sample $l$ and the phase-mismatching wave vector $\Delta\vec{k}=\vec{k}_{A}-\vec{k}_{0}-\vec{k}_{S}$ are required to satisfy $\vert\Delta\vec{k}\vert l\ll2\pi$~\cite{Chen2020Enantio,Chen2021arXivEnantio,lou2014Competition}. In addition, for simplicity, we have ignored the couplings among electronic, vibrational, and rotational degrees of freedom~\cite{shapiro2000Coherently,gerbasi2001Theory,brumer2001Principles,frishman2004Optical}. Note that enantio-discrimination~\cite{Patterson2013Enantiomer,Patterson2013Sensitive,
Patterson2014New,Shubert2014Identifying,Shubert2015Rotational,Shubert2016Chiral,Lobsiger2015Molecular} and enantio-specific state transfer~\cite{Eibenberger2017Enantiomer,Perez2017Coherent,Lee2022Quantitative} have been experimentally achieved by using the three-level $\Delta$-type model of chiral molecules. This three-level model offers possible experimental techniques for implementing the current four-level chiral-molecule systems.

Under the dipole approximation and rotating-wave approximation, the Hamiltonian of the system reads ($\hbar=1$)
\begin{eqnarray}\label{Hs}
\hat{H}_{s}&=&\omega_{S}\vert S\rangle\langle S\vert+\omega_{A}\vert A\rangle\langle A\vert+(\eta\vert L\rangle\langle R\vert+\Omega_{0}e^{i\omega_{0}t}\vert S\rangle\langle A\vert+\mathrm{H.c.})\nonumber\\
&&+\sum_{Q=L,R}(\Omega_{S}e^{i\omega_{1}t}\vert Q\rangle\langle S\vert+\Omega_{A}e^{i\phi_{Q}}e^{i\omega_{2}t}\vert Q\rangle\langle A\vert+\mathrm{H.c.}),
\end{eqnarray}
where $\eta$ denotes the tunneling strength between the two chiral ground states $\vert L\rangle$ and $\vert R\rangle$. The parameters $\Omega_{0}$, $\Omega_{S}$, and $\Omega_{A}$ ($\omega_{0}$, $\omega_{1}$, and $\omega_{2}$) are, respectively, the coupling strengths (frequencies) of three electromagnetic fields applied to different electric-dipole transitions $\vert S\rangle\leftrightarrow\vert A\rangle$, $\vert Q\rangle\leftrightarrow\vert S\rangle$, and $\vert Q\rangle\leftrightarrow\vert A\rangle$ for $Q=L,R$. $\phi_{Q}$ ($Q=L,R$) is the overall phase of the $\Delta$-type substructure $\vert Q\rangle\leftrightarrow\vert S\rangle\leftrightarrow\vert A\rangle\leftrightarrow\vert Q\rangle$. The molecular chirality is reflected in the overall phases of the two $\Delta$-type substructures, i.e., $\phi_{L}=\phi$ and $\phi_{R}=\phi+\pi$.

We are interested in the case of the one-photon resonance of $\vert Q\rangle\leftrightarrow\vert A\rangle$ and the three-photon resonance, i.e.,
\begin{equation}
\omega_{2}=\omega_{A},\quad \omega_{A}=\omega_{1}+\omega_{0}.
\end{equation}
In the interaction picture with respect to $\hat{\mathcal{H}}_{0}=\omega_{1}\vert S\rangle\langle S\vert+\omega_{2}\vert A\rangle\langle A\vert$, the Hamiltonian~(\ref{Hs}) becomes
\begin{eqnarray}
\hat{H}& =& e^{i\hat{\mathcal{H}}_{0}t}(\hat{H}_{s}-\hat{\mathcal{H}}_{0})e^{-i\hat{\mathcal{H}}_{0}t}\nonumber\\
& =&\Delta\vert S\rangle\langle S\vert+(\eta\vert L\rangle\langle R\vert+\Omega_{0}\vert S\rangle\langle A\vert+\mathrm{H.c.})\nonumber\\
&&+\sum_{Q=L,R}(\Omega_{S}\vert Q\rangle\langle S\vert+\Omega_{A}e^{i\phi_{Q}}\vert Q\rangle\langle A\vert+\mathrm{H.c.}),\label{Ham}
\end{eqnarray}
where $\Delta=\omega_{S}-\omega_{1}$ is the detuning between the transition $\vert S\rangle\leftrightarrow\vert Q\rangle$ and the applied driving field of frequency $\omega_{1}$. For simplicity, we have assumed that $\eta$, $\Omega_{0}$, $\Omega_{S}$, and $\Omega_{A}$ are real.

Under the condition of large detuning $\vert\Delta\vert\gg\Omega_{S}\sim\Omega_{0}\gg\Omega_{A}\sim \eta$, the effective Hamiltonian can be obtained by using the Fr\"{o}hlich-Nakajima transformation~\cite{Frohlich1950Theory,Nakajima1955Perturbation} to eliminate adiabatically the symmetric achiral excited state $\vert S\rangle$. For that, we introduce an anti-Hermitian operator
\begin{equation}
\hat{\mathcal{S}}=\frac{1}{\Delta}\left[\Omega_{S}\left(\vert L\rangle+\vert R\rangle\right)\langle S\vert+\Omega_{0}\vert A\rangle\langle S\vert-\mathrm{H.c.}\right],
\end{equation}
which is determined by the equation $[\hat{H}_{0},\hat{\mathcal{S}}]+\hat{H}_{1}=0$. Here the zero-, first-, and second-order Hamiltonians are, respectively, $\hat{H}_{0}=\Delta\vert S\rangle\langle S\vert$, $\hat{H}_{1}=\Omega_{S}\left(\vert L\rangle+\vert R\rangle\right)\langle S\vert+\Omega_{0}\vert S\rangle\langle A\vert+\mathrm{H.c.}$, and $\hat{H}_{2} =(\sum_{Q=L,R}\Omega_{A}e^{i\phi_{Q}}\vert Q\rangle\langle A\vert+\eta\vert L\rangle\langle R\vert+\mathrm{H.c.})$.

Up to the second order, the effective Hamiltonian can be obtained as~\cite{Li2007Generalized,Frohlich1950Theory,Nakajima1955Perturbation,Li2007Time}
\begin{eqnarray}\label{effHam}
\hat{H}_{\mathrm{eff}} & =&\exp(-\hat{\mathcal{S}})\hat{H}\exp(\hat{\mathcal{S}})\simeq \hat{H}_{0}+[\hat{H}_{1},\hat{\mathcal{S}}]/2+\hat{H}_{2}\nonumber \\
& =&\tilde{\Delta}\vert S\rangle\langle S\vert+\tilde{\Lambda}\vert A\rangle\langle A\vert+\Lambda(\vert L\rangle\langle L\vert+\vert R\rangle\langle R\vert)\nonumber \\
&&+[\sum_{Q=L,R}\tilde{\Omega}_{Q}\vert Q\rangle\langle A\vert+(\eta+\Lambda)\vert L\rangle\langle R\vert+\mathrm{H.c.}],
\end{eqnarray}
where we have defined $\Lambda\equiv-\Omega_{S}^{2}/\Delta$, $\tilde{\Lambda}\equiv-\Omega_{0}^{2}/\Delta$, $\tilde{\Delta}\equiv\Delta-2\Lambda-\tilde{\Lambda}$, and
\begin{equation}
\tilde{\Omega}_{Q}\equiv\Omega_{A}e^{i\phi_{Q}}-\frac{\Omega_{S}\Omega_{0}}{\Delta}.
\end{equation}
It can be seen from Eq.~(\ref{effHam}) that the symmetric achiral excited states $\vert S\rangle$ is decoupled to other three states ($\vert L\rangle$, $\vert R\rangle$, and $\vert A\rangle$), then the evolution of two chiral ground states $\vert Q\rangle$ will not be affected by $\vert S\rangle$. Hence, the dynamics of the system can be described by the following reduced three-level Hamiltonian
\begin{eqnarray}\label{reeffHam}
\hat{H}_{\mathrm{re}} &=&\tilde{\Lambda}\vert A\rangle\langle A\vert+\Lambda(\vert L\rangle\langle L\vert+\vert R\rangle\langle R\vert)\nonumber\\
&&+[\sum_{Q=L,R}\tilde{\Omega}_{Q}\vert Q\rangle\langle A\vert+(\eta+\Lambda)\vert L\rangle\langle R\vert+\mathrm{H.c.}].
\end{eqnarray}
Figure~\ref{Fig1}(b) shows this effective three-level model. Here the blue solid and dashed lines express the single photon processes $\vert Q\rangle\leftrightarrow\vert A\rangle$ and the two-photon processes $\vert Q\rangle\leftrightarrow\vert S\rangle\leftrightarrow\vert A\rangle$, respectively. The orange solid (dashed) line corresponds to the direct (indirect) tunneling interaction between the two chiral ground states. This indirect tunneling interaction with coupling strength $\Lambda\equiv-\Omega_{S}^{2}/\Delta$ is induced by the process $\vert L\rangle\leftrightarrow\vert S\rangle\leftrightarrow\vert R\rangle$.

In order to establish the chiral-state-selective excitations with the right-handed chiral ground state being excited to the asymmetric achiral excited state and the left-handed one being undisturbed, the appropriate detuning and coupling strengths of the electromagnetic fields should be chosen to eliminate the tunneling interaction between two chiral ground states and the interaction between $\vert L\rangle$ and $\vert A\rangle$. For this purpose, we design the system parameters to satisfy
\begin{equation}
\Delta=\frac{\Omega_{S}^{2}}{\eta}\equiv\Delta_{0},\quad
\phi=0,\quad
\Omega_{A}=\frac{\Omega_{S}\Omega_{0}}{\Delta},
\end{equation}
i.e., $\eta+\Lambda=0$, $\tilde{\Omega}_{L}=0$, and $\tilde{\Omega}_{R}=-2\Omega_{A}$. Under these conditions, the reduced three-level Hamiltonian can be written as
\begin{equation}\label{reeffHams}
\hat{H}_{\mathrm{re}} =\tilde{\Lambda}\vert A\rangle\langle A\vert+\sum_{Q=L,R}\Lambda\vert Q\rangle\langle Q\vert-2\Omega_{A}(\vert R\rangle\langle A\vert+\mathrm{H.c.}).
\end{equation}
Here the right-handed chiral ground state $\vert R\rangle$ is coupled to the asymmetric achiral excited state $\vert A\rangle$ and the left-handed chiral ground state $\vert L\rangle$ is decoupled to $\vert R\rangle$ and $\vert A\rangle$ [see Fig.~\ref{Fig1}(c)], indicating that the chiral-state-selective excitations can be realized. Note that when the overall phase $\phi=0$ changes to be $\phi=\pi$, one has $\tilde{\Omega}_{L}=-2\Omega_{A}$ and $\tilde{\Omega}_{R}=0$. This means that $\vert L\rangle$ is coupled to $\vert A\rangle$, and $\vert R\rangle$ is decoupled to $\vert L\rangle$ and $\vert A\rangle$.

\section{Chiral-state-selective excitations\label{Selexcsec}}
In this section, we will demonstrate the chiral-state-selective excitations in the absence of the system dissipation by calculating the populations of two chiral ground states. We consider an initial racemic mixture, namely, the initial state of each molecule can be described by density operator $\rho(0)=(\vert L\rangle\langle L\vert+\vert R\rangle\langle R\vert)/2$. By numerically solving the Liouville equation $d\rho/dt=-i[\hat{H},\rho]$ [$\hat{H}$ is given in Eq.~(\ref{Ham})], we can obtain the density operator $\rho(t)$ of the system at time $t$ and the evolved populations of the left- and right-handed chiral ground states $P_{Q}(t)=\langle Q\vert\rho(t)\vert Q\rangle$.
\begin{figure}
\center
\includegraphics[width=0.6 \textwidth]{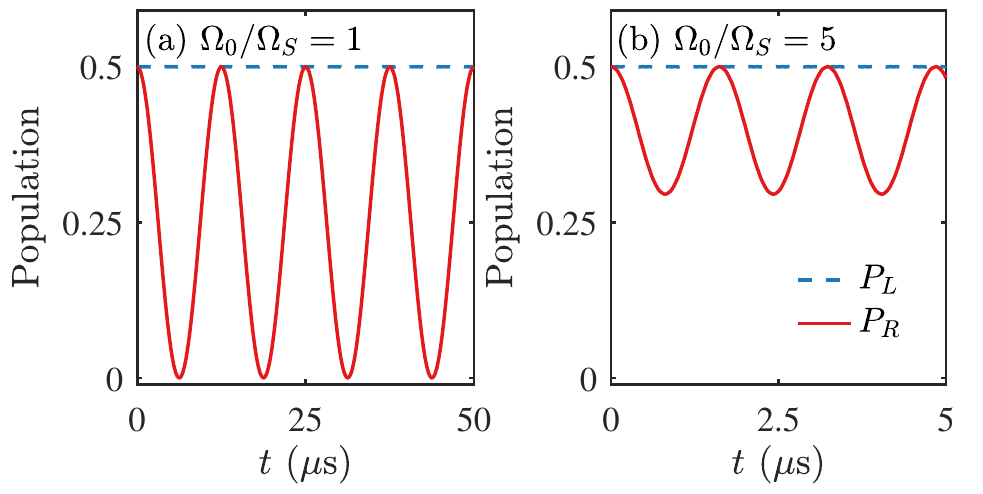}
\caption{The evolved populations of the left- and right-handed chiral ground states in the absence of the system dissipation at (a) $\Omega_{0}/\Omega_{S}=1$ and (b) $\Omega_{0}/\Omega_{S}=5$. Here the dashed and solid curves correspond to populations of the left- and right-handed chiral ground states, respectively. Other parameters are $\phi=0$, $\eta/2\pi=0.02\,\mathrm{MHz}$, $\Omega_{S}/2\pi=1\,\mathrm{MHz}$, $\Delta/2\pi=50\,\mathrm{MHz}$, and $\Omega_{A}=\Omega_{S}\Omega_{0}/\Delta$.}
\label{Fig2}
\end{figure}

The results showing the time evolution of the populations $P_{Q}$ ($Q=L,R$) of the left- and right-handed chiral ground states at the coupling strength $\Omega_{0}/\Omega_{S}=1$ are presented in Fig.~\ref{Fig2}(a). Other parameters are $\phi=0$, $\eta/2\pi=0.02\,\mathrm{MHz}$~\cite{thanopulos2003Theory}, $\Omega_{S}/2\pi=1\,\mathrm{MHz}$ (which mean $\Delta_{0}\equiv\Omega_{S}^{2}/\eta=2\pi\times50\,\mathrm{MHz}$), $\Delta/2\pi=50\,\mathrm{MHz}$, and $\Omega_{A}=\Omega_{S}\Omega_{0}/\Delta=2\pi\times0.02\,\mathrm{MHz}$, which are experimentally feasible~\cite{Patterson2013Enantiomer,Patterson2013Sensitive,Eibenberger2017Enantiomer,Perez2017Coherent,Lee2022Quantitative}. We can see that, the evolved population $P_{L}$ is almost unchanged and the population $P_{R}$ appears as a periodic oscillation. This indicates that the left-handed chiral ground state $\vert L\rangle$ is almost undisturbed and the right-handed one $\vert R\rangle$ can be excited to an achiral excited state, i.e., establishing the chiral-state-selective excitations. Figure~\ref{Fig2}(b) depicts the evolved populations $P_{Q}$ ($Q=L,R$) at $\Omega_{0}/\Omega_{S}=5$, where $\Omega_{A}=\Omega_{S}\Omega_{0}/\Delta=2\pi\times0.1\,\mathrm{MHz}$. Similarly, it is shown that the population $P_{L}$ is almost unchanged and the population $P_{R}$ appears as a periodic oscillation. In particular, we find that, comparing with that in Fig.~\ref{Fig2}(a), the period of oscillation for the population $P_{R}$ becomes shorter and its amplitude becomes smaller when the coupling strength increases to $\Omega_{0}/\Omega_{S}=5$ in Fig.~\ref{Fig2}(b). The reasons are the following: (i) The period of oscillation for the population $P_{R}$ is dependent on the coupling strength $2\Omega_{A}$ between $\vert R\rangle$ and $\vert A\rangle$. When $\Omega_{0}/\Omega_{S}=1$ ($\Omega_{0}/\Omega_{S}=5$), the coupling strength is $2\Omega_{A}=2\pi\times0.04\,\mathrm{MHz}$ ($2\Omega_{A}=2\pi\times0.2\,\mathrm{MHz}$). Hence, compared with $\Omega_{0}/\Omega_{S}=1$, the period of oscillation becomes shorter when $\Omega_{0}/\Omega_{S}=5$. (ii) The amplitude of oscillation for the population $P_{R}$ depends on the detuning $\delta\equiv\Lambda-\tilde{\Lambda}$ for the interaction term $-2\Omega_{A}[\vert R\rangle\langle A\vert\exp{(i\delta t)}+\mathrm{H.c.}]$ of Hamiltonian~(\ref{reeffHams}) in the interaction picture. At $\Omega_{0}/\Omega_{S}=1$, we find $\delta\equiv\Lambda-\tilde{\Lambda}=0$, i.e., the resonance coupling between $\vert R\rangle$ and $\vert A\rangle$ occurs. However, we find that the detuning $\delta\neq0$ when $\Omega_{0}/\Omega_{S}=5$, thus the amplitude of oscillation is decreased compared with the case $\Omega_{0}/\Omega_{S}=1$.

\section{Enantio-conversion via optical pumping\label{Enconsec}}

In Sec.~\ref{Selexcsec}, we have discussed the chiral-state-selective excitations in the absence of the system dissipation. However, in the realistic situations, the system dissipation is inevitable and crucial for implementing enantio-conversion of chiral mixtures. In the following, we will demonstrate that high-efficiency enantio-conversion of chiral mixtures via optical pumping can be realized under the action of the combining effect of the system dissipation and the chiral-state-selective excitations.

The dynamics of the system is governed by the quantum master equation
\begin{equation}\label{QME}
\frac{d\rho}{dt}=-i[\hat{H},\rho]+\mathcal{L}\rho,
\end{equation}
where $\hat{H}$ is given in Eq.~(\ref{Ham}) and $\mathcal{L}\rho\equiv\mathcal{L}_{\mathrm{dc}}\rho+\mathcal{L}_{\mathrm{dp}}\rho$ is the Lindblad superoperator that describes the dissipation of the system. The decay term $\mathcal{L}_{\mathrm{dc}}\rho$ reads~\cite{scully1997Quantum}
\begin{equation}\label{determ}
\mathcal{L}_{\mathrm{dc}}\rho=\frac{\gamma_{S}}{2}(\mathcal{L}_{\hat{\sigma}_{LS}}\rho+\mathcal{L}_{\hat{\sigma}_{RS}}\rho)
+\frac{\gamma_{SA}}{2}\mathcal{L}_{\hat{\sigma}_{SA}}\rho+\frac{\gamma_{A}}{2}(\mathcal{L}_{\hat{\sigma}_{LA}}\rho+\mathcal{L}_{\hat{\sigma}_{RA}}\rho),
\end{equation}
where $\gamma_{S}$ ($\gamma_{A}$) is the chirality-independent~\cite{frishman2004Optical} decay rate of the chiral molecules from state $\vert S\rangle$ ($\vert A\rangle$) to $\vert Q\rangle$ for $Q=L,R$ and $\gamma_{SA}$ is the decay rate from state $\vert A\rangle$ to $\vert S\rangle$. In addition, we have defined that the operator $\hat{\sigma}_{pq}=\vert p\rangle\langle q\vert$ with $p,q=L,R,S,A$ and the
Lindblad superoperator $\mathcal{L}_{\hat{o}}\rho=2\hat{o}\rho\hat{o}^{\dagger}-\hat{o}^{\dagger}\hat{o}\rho-\rho\hat{o}^{\dagger}\hat{o}$ with $\hat{o}=\hat{\sigma}_{LS},\hat{\sigma}_{RS},\hat{\sigma}_{LA},\hat{\sigma}_{RA},\hat{\sigma}_{SA}$. The pure dephasing term $\mathcal{L}_{\mathrm{dp}}\rho$ reads~\cite{Kang2020Effective,Hauss2008Single}
\begin{equation}\label{dpterm}
\mathcal{L}_{\mathrm{dp}}\rho=\frac{\gamma_{\phi}}{2}[\sum_{Q=L,R}(\mathcal{L}_{\hat{\sigma}^{z}_{SQ}}\rho+\mathcal{L}_{\hat{\sigma}^{z}_{AQ}}\rho)
+\mathcal{L}_{\hat{\sigma}^{z}_{AS}}\rho+\mathcal{L}_{\hat{\sigma}^{z}_{RL}}\rho],
\end{equation}
where we have defined the operator $\hat{\sigma}^{z}_{pq}=\vert p\rangle\langle p\vert-\vert q\rangle\langle q\vert$. For simplicity, we have assumed that the pure dephasing rate $\gamma_{\phi}$ is state-independent~\cite{frishman2004Optical,Patterson2012Cooling}. Similarly, we consider an initial state $\rho(0)=(\vert L\rangle\langle L\vert+\vert R\rangle\langle R\vert)/2$ of the system, i.e., $P_{L}(0)=P_{R}(0)=1/2$. By numerically solving Eq.~(\ref{QME}), we can obtain the evolved density operator $\rho(t)$ of the system and the evolved populations of two chiral ground states $P_{Q}(t)=\langle Q\vert\rho(t)\vert Q\rangle$. In this work, all numerical computations are performed by using QuTip (Quantum Toolbox in Python)~\cite{johansson2012QuTiP,johansson2013QuTiP}.
\begin{figure}
\center
\includegraphics[width=0.6 \textwidth]{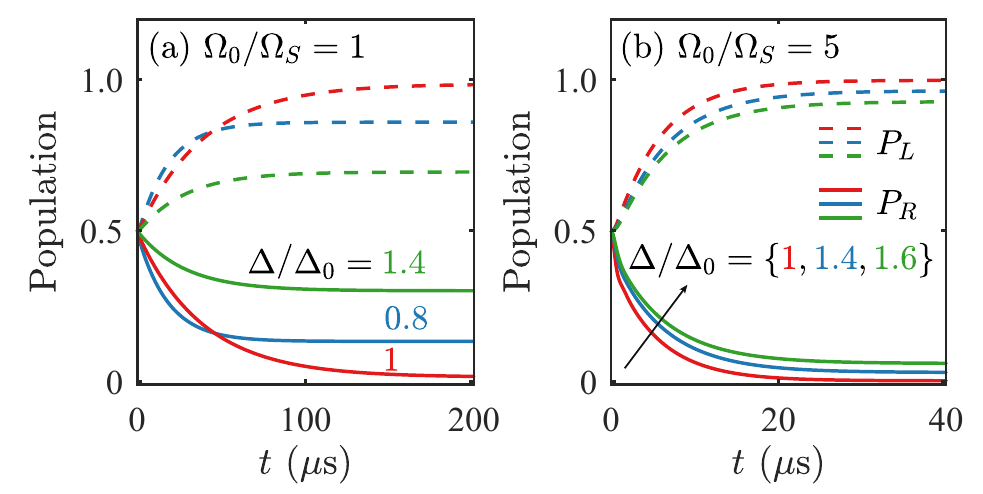}
\caption{The evolved populations of the left- and right-handed chiral ground states in the presence of the system dissipation for different detunings $\Delta$ at (a) $\Omega_{0}/\Omega_{S}=1$ and (b) $\Omega_{0}/\Omega_{S}=5$. Here the dashed and solid curves correspond to populations of the left- and right-handed chiral ground states, respectively. Other parameters are $\phi=0$, $\eta/2\pi=0.02\,\mathrm{MHz}$, $\Omega_{S}/2\pi=1\,\mathrm{MHz}$ (correspondingly $\Delta_{0}/2\pi=50\,\mathrm{MHz}$), $\Omega_{A}=\Omega_{S}\Omega_{0}/\Delta$, $\gamma_{S}/2\pi=\gamma_{A}/2\pi=0.1\,\mathrm{MHz}$, $\gamma_{SA}/2\pi=0.5\,\mathrm{MHz}$, and $\gamma_{\phi}/2\pi=0.01\,\mathrm{MHz}$.}
\label{Fig3}
\end{figure}

In order to analyze the efficiency of the enantio-conversion of the chiral mixtures, we show the time evolution of the populations of the left- and right-handed chiral ground states in the presence of the system dissipation. Specifically, we choose the experimentally feasible decay rates~\cite{Patterson2013Enantiomer,Patterson2013Sensitive}: $\gamma_{S}/2\pi=\gamma_{A}/2\pi=0.1\,\mathrm{MHz}$, $\gamma_{SA}/2\pi=0.5\,\mathrm{MHz}$, and $\gamma_{\phi}/2\pi=0.01\,\mathrm{MHz}$. In Fig.~\ref{Fig3}(a), we show the evolved populations $P_{Q}(t)$ for different detunings $\Delta$ at the coupling strength $\Omega_{0}/\Omega_{S}=1$. It can be seen that the population of the left-handed (right-handed) chiral ground state is approximately equal to 1 (0) at the detuning $\Delta/\Delta_{0}=1$ when $t>150\,\mu$s, which indicates that the conversion of the chiral mixtures from $\vert R\rangle$ to $\vert L\rangle$ is achieved under the action of the combining effect of the system dissipation and the chiral-state-selective excitations. In particular, we calculate the enantiomeric excess $\varepsilon\equiv( P_{L}-P_{R})/(P_{L}+P_{R})$ of the chiral ground state to estimate the efficiency of the enantio-conversion of the chiral mixtures. From Fig.~\ref{Fig3}(a), we find that the steady-state enantiomeric excess is $\varepsilon\approx98.3\%$ at $\Delta/\Delta_{0}=1$ when $t>150\,\mu$s, i.e., achieving high-efficiency enantio-conversion of the chiral mixtures. In addition, we find that the population $P_{L}$ ($P_{R}$) decreases (increases) when the detuning $\Delta$ diverges from $\Delta_{0}$, meaning that the enantiomeric excess is decreased, i.e., the efficiency of the enantio-conversion of the chiral mixtures decreases when the detuning $\Delta$ diverges from $\Delta_{0}$. The reason is that the tunneling interaction between two chiral ground states can be eliminated only when $\Delta=\Delta_{0}$. In Fig.~\ref{Fig3}(b), we depict the evolved populations $P_{Q}$ for different detunings at $\Omega_{0}/\Omega_{S}=5$. It is shown that the populations are $P_{L}\approx0.998$ and $P_{R}\approx0.001$ (i.e., the steady-state enantiomeric excess $\varepsilon\approx99.8\%$) at $\Delta/\Delta_{0}=1$ when $t>20\,\mu$s, which indicates that the high-efficiency enantio-conversion of the chiral mixtures can be achieved and the required time to complete enantio-conversion is shorter. In addition, we find that the population $P_{L}$ ($P_{R}$) decreases (increases) slowly as the detuning $\Delta$ deviates from $\Delta_{0}$. Note that once the three electromagnetic fields are turned off, the system will begin to oscillate between $\vert L\rangle$ and $\vert R\rangle$ as a result of tunneling.
\begin{figure}
\center
\includegraphics[width=0.6 \textwidth]{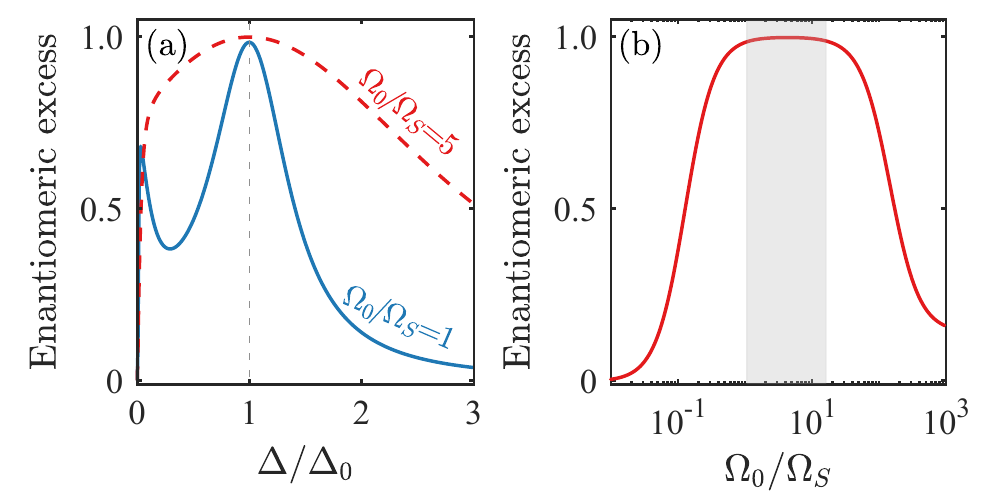}
\caption{(a) The steady-state enantiomeric excess $\varepsilon$ of the chiral ground state as a function of the detuning $\Delta$ when the coupling strength $\Omega_{0}/\Omega_{S}=1$ and $5$. (b) The steady-state enantiomeric excess $\varepsilon$ as a function of the coupling strength $\Omega_{0}$ at $\Delta/\Delta_{0}=1$. Other parameters are the same as those in Fig.~\ref{Fig3}.}
\label{Fig4}
\end{figure}

To demonstrate the effect of the detuning $\Delta$ on enantio-conversion of chiral mixtures, we show the steady-state enantiomeric excess $\varepsilon$ of the chiral ground state versus the detuning $\Delta$ when the coupling strength $\Omega_{0}/\Omega_{S}=1$ and $5$ in Fig.~\ref{Fig4}(a). It can be seen that the steady-state enantiomeric excess (i.e., $\varepsilon\approx 98.3\%$ or $99.8\%$) is obtained at $\Delta/\Delta_{0}=1$ when $\Omega_{0}/\Omega_{S}=1$ or $5$, namely, the high-efficiency enantio-conversion of chiral mixtures is realized. In addition, we find that, at $\Omega_{0}/\Omega_{S}=1$ ($\Omega_{0}/\Omega_{S}=5$), the enantiomeric excess decreases quickly (slowly) as the detuning $\Delta$ deviates from $\Delta_{0}$. In Fig.~\ref{Fig4}(b), the steady-state enantiomeric excess $\varepsilon$ is plotted versus the coupling strength $\Omega_{0}$ at $\Delta/\Delta_{0}=1$. We observe that the enantiomeric excess $\varepsilon$ increases (decreases) with the increase of the coupling strength $\Omega_{0}$ when $\Omega_{0}/\Omega_{S}<1$ ($\Omega_{0}/\Omega_{S}>20$). In a middle region of $1<\Omega_{0}/\Omega_{S}<20$, the enantiomeric excess $\varepsilon$ can reach maximum, which means that the high-efficiency enantio-conversion of chiral mixtures can be realized in this region at $\Delta/\Delta_{0}=1$. The reason is that in this middle region, the parameters of the system are closer to the condition $\vert\Delta\vert\gg\Omega_{S}\sim\Omega_{0}\gg\Omega_{A}\sim \eta$ of the adiabatic elimination.

\begin{figure}
\center
\includegraphics[width=0.6 \textwidth]{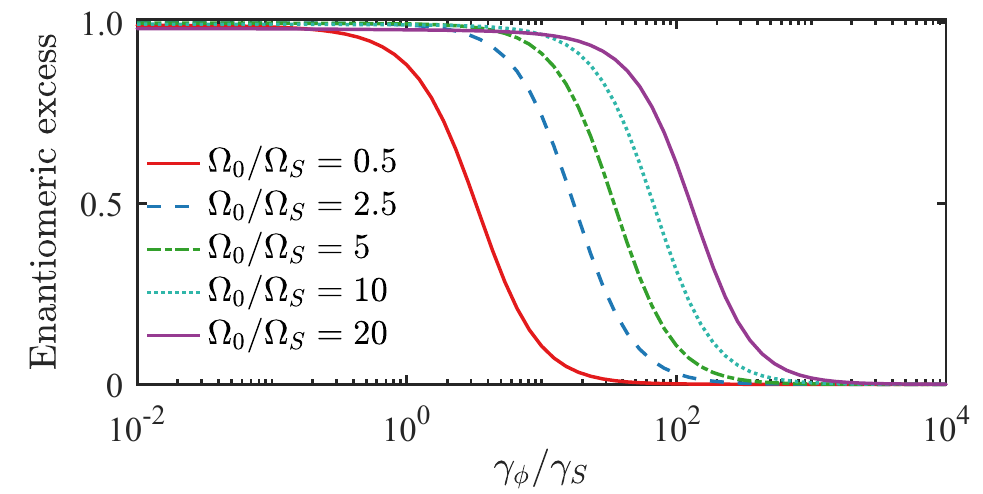}
\caption{The steady-state enantiomeric excess $\varepsilon$ of the chiral ground state as a function of the dephasing rate $\gamma_{\phi}$ for different coupling strengths $\Omega_{0}$. Other parameters are the same as those in Fig.~\ref{Fig3} expect
for $\Omega_{S}/2\pi=2\,\mathrm{MHz}$ (correspondingly $\Delta_{0}/2\pi=200\,\mathrm{MHz}$) and $\Delta/2\pi=200\,\mathrm{MHz}$.}
\label{Fig5}
\end{figure}

We also analyze how the efficiency of the enantio-conversion in the chiral mixtures depends on the dephasing rate. Specifically, we show the steady-state enantiomeric excess $\varepsilon$ of the chiral ground state as a function of the dephasing rate $\gamma_{\phi}$ for different coupling strengths $\Omega_{0}$ in Fig.~\ref{Fig5}. It can be seen that the steady-state enantiomeric excess $\varepsilon$ decreases with the increase of the dephasing rate $\gamma_{\phi}$. This means that the dephasing will affect evidently the enantio-conversion of chiral mixtures. In addition, we find that as the coupling strength $\Omega_{0}$ increases, it becomes possible to achieve the high-efficiency enantio-conversion of chiral mixtures, even in cases where the dephasing rate is larger than the other decay rates. For example, for the large-dephasing case of $\gamma_{\phi}/\gamma_{S}=30$, the steady-state enantiomeric excess $\varepsilon$ is about $91.6\%$ when $\Omega_{0}/\Omega_{S}=20$ (purple solid curve).

\begin{figure}
\center
\includegraphics[width=0.6 \textwidth]{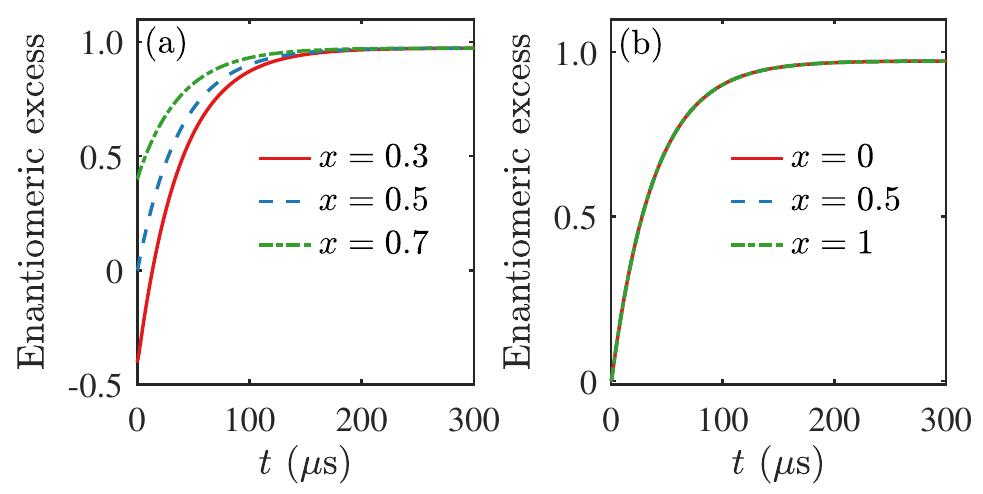}
\caption{The time evolution of the enantiomeric excess $\varepsilon$ of the chiral ground state when the initial state of each molecule (a) $\rho(0)=x\vert L\rangle\langle L\vert+(1-x)\vert R\rangle\langle R\vert$ and (b) $\rho(0)=x\vert +\rangle\langle +\vert+(1-x)\vert -\rangle\langle -\vert$. Here $x$ is the coefficient and $\vert \pm\rangle=(\vert L\rangle\pm\vert R\rangle)/\sqrt{2}$ are the symmetric and antisymmetric tunneling-splitting eigenstates. Other parameters are $\phi=0$, $\eta/2\pi=0.02\,\mathrm{MHz}$, $\Omega_{0}/2\pi=\Omega_{S}/2\pi=1\,\mathrm{MHz}$, $\Delta/2\pi=50\,\mathrm{MHz}$, $\Omega_{A}=\Omega_{S}\Omega_{0}/\Delta$, $\gamma_{S}/2\pi=\gamma_{A}/2\pi=0.1\,\mathrm{MHz}$, $\gamma_{SA}/2\pi=0.5\,\mathrm{MHz}$, and $\gamma_{\phi}/2\pi=0.01\,\mathrm{MHz}$.}
\label{Fig6}
\end{figure}

In the above discussions, we have mainly focused on the case that the initial chiral mixture is a racemic mixture, i.e., the initial state of each molecule is $\rho(0)=(\vert L\rangle\langle L\vert+\vert R\rangle\langle R\vert)/2$. Below, we will discuss the influence of different initial states of the system on the enantiomeric excess of the chiral ground state. In Fig.~\ref{Fig6}(a), we show the time evolution of the enantiomeric excess when the initial state of each molecule is $\rho(0)=x\vert L\rangle\langle L\vert+(1-x)\vert R\rangle\langle R\vert$ with $x=0.3, 0.5, 0.7$. It can be seen that the steady-state enantiomeric excess is the same for the different initial conditions. Figure~\ref{Fig6}(b) shows the time evolution of the enantiomeric excess when the initial state of each molecule is $\rho(0)=x\vert +\rangle\langle +\vert+(1-x)\vert -\rangle\langle -\vert$ with $x=0, 0.5, 1$. Here $\vert \pm\rangle=(\vert L\rangle\pm\vert R\rangle)/\sqrt{2}$ are the symmetric and antisymmetric tunneling-splitting eigenstates. The results demonstrate that for the initial states $\rho(0)=x\vert +\rangle\langle +\vert+(1-x)\vert -\rangle\langle -\vert$ with $x=0, 0.5$, and $1$, the time evolution of the enantiomeric excess is identical in all three cases.

\section{Conclusion \label{conclusion}}
Based on the four-level model of the chiral molecules, we have demonstrated that high-efficiency enantio-conversion of chiral mixtures can be achieved via optical pumping when the tunneling interaction between two chiral ground states can not be ignored. In this four-level model, the chiral-state-selective excitations can be established by choosing the appropriate detuning and coupling strengths of the electromagnetic fields. The numerical results show that, under the action of the combining effect of the system dissipation and the chiral-state-selective excitations, high-efficiency enantio-conversion of chiral mixtures can be achieved when $\Delta=\Omega_{S}^{2}/\eta$ and $\Omega_{A}=\Omega_{S}\Omega_{0}/\Delta$ in the large-detuning region. In addition, by analyzing the dependence of the enantiomeric excess on the detuning and coupling strengths of the electromagnetic fields, the optimal parameters of achieving high-efficiency enantio-conversion of chiral mixtures have been analyzed. Our work opens up a route to achieve high-efficiency enantio-conversion of chiral mixtures in the presence of the tunneling interaction between two chiral states.

\section*{Acknowledgments}
This work is supported in part by the National Natural Science Foundation of China (Grants No.~12074030, No.~12274107, No.~12105011, No.~U2230402, and No.~12147109), the China Postdoctoral Science Foundation (Grant No.~2021M700360), and the Research Funds of Hainan University [Grant No.~KYQD(ZR)23010].

\section*{Data availability statement}
All data that support the findings of this study are included within the article (and any supplementary files).

\section*{References}

\providecommand{\newblock}{}

\end{document}